\documentclass[12pt]{article}
\usepackage{amsmath,amssymb,latexsym}

\newcommand{\beq}{\begin{equation}}
\newcommand{\eeq}{\end{equation}}
\newcommand{\bea}{\begin{eqnarray}}
\newcommand{\eea}{\end{eqnarray}}
\newcommand{\beqar}{\begin{eqnarray*}}
\newcommand{\eeqar}{\end{eqnarray*}}

\begin{document}

\newpage
\bigskip
\hskip 5in\vbox{\baselineskip12pt
\hbox{hep-th/0008211}
\hbox{UK/00-05}
\hbox{DTP/00/75}}

\bigskip
\bigskip
\bigskip

\centerline{\bf \Large Higher Dimensional Kerr--AdS Black Holes}
\medskip 
\centerline{\bf \Large and the}
\medskip
 \centerline{\bf\Large AdS/CFT
  Correspondence}

\bigskip
\bigskip
\bigskip

\centerline{ Adel M. Awad{$^\sharp$} 
and Clifford V. Johnson$^\natural$}
\bigskip
\bigskip
\bigskip

\centerline{\it $^\sharp$Department of Physics and Astronomy,
University of Kentucky,  Lexington, KY 40506, U.S.A.}
\centerline{\it and}
\centerline{\it Department of Physics, 
Faculty of Science, Ain Shams University, Cairo 11566, Egypt}
\bigskip
\bigskip
\centerline{\it $^\natural$Centre 
for Particle Theory, Department of Mathematical Sciences}

\centerline{\it University of
Durham, Durham, DH1 3LE, U.K.}

\bigskip

\centerline{\tt $^\sharp$adel@pa.uky.edu, $^\natural$c.v.johnson@durham.ac.uk}

\bigskip

\begin{abstract} 
  \medskip Using the counterterm subtraction technique we calculate
  the stress--energy tensor, action, and other physical quantities for
  Kerr--AdS black holes in various dimensions.  For Kerr--AdS$_5$ with
  both rotation parameters non--zero, we demonstrate that
  stress--energy tensor, in the zero mass parameter limit, is equal
  to the stress tensor of the weakly coupled four dimensional dual
  field theory.  As a result, the total energy of the general
  Kerr--AdS$_5$ black hole at zero mass parameter, exactly matches the
  Casimir energy of the dual field theory. We show that at high
  temperature, the general Kerr--AdS$_5$ and perturbative field theory
  stress--energy tensors are equal, up to the usual factor of 3/4. We
  also use the counterterm technique to calculate the stress tensors
  and actions for Kerr--AdS$_{6}$, and Kerr--AdS$_{7}$ black holes,
  with one rotation parameter, and we display the results.  We discuss
  the conformal anomalies of the field theories dual to the
  Kerr--AdS$_{5}$ and Kerr--AdS$_{7}$ spacetimes. In these two field
  theories, we show that the rotation parameters break conformal
  invariance but not scale invariance, a novel result for a
  non--trivial field theory. For Kerr--AdS$_7$ the conformal anomalies
  calculated on the gravity side and the dual (0,2) tensor multiplet
  theory are equal up to 4/7 factor. We expect that the Casimir energy
  of the free field theory is the same as the energy of the
  Kerr--AdS$_7$ black hole (with zero mass parameter), up to that
  factor. \end{abstract} \newpage \baselineskip=18pt
\setcounter{footnote}{0}

\section{Introduction} 
There has been considerable interest in Anti de--Sitter (AdS) black
holes since the discovery of the AdS/CFT
correspondence\cite{maldacena}, relating the physics of AdS in $n+1$
dimensions to a strongly coupled conformal (gauge) field theory (CFT)
in one dimension fewer.  This interest was triggered by Witten
\cite{witten}, who in formulating a precise statement of the duality,
(see also ref.\cite{gubklebpoly}) found the dual interpretation of the
thermodynamical properties of Schwarzchild AdS black holes (Sch--AdS)
in terms of the dual field theory at a non--zero temperature related
to the mass of the black hole.  An example of this remarkable
connections is the interpretation of the Hawking--Page phase
transition\cite{hawkpage} from AdS to Sch--AdS at finite temperature,
as terms of a transition between phases of the dual strongly coupled
field theory.

Since then many new works have been presented on AdS black holes with
the three simplest properties of mass, charge and rotation, in the
context of the AdS/CFT correspondence.  Reissner--Nordstrom AdS
(RN--AdS) black holes were studied in this context in
ref.\cite{clifford}. There are again remarkable interpretations of the
black holes' properties in terms of the dual gauge field theory. There
is a beautiful family of phase transitions as a function of
temperature and charge, (also temperature and potential). These are
interpreted in the gauge theory as a family of transitions at finite
temperature and a chemical potential for a global charge under the
R--symmetry of the (broken) supersymmetry of the gauge theory.  This
latter is especially interesting, since for the appropriate choice of
R--current, the chemical potential in the dual gauge theory would be a
close cousin of the baryon density operator in QCD.  Therefore quite
remarkably, studies of such charged black holes might give a window on
the phase structure of gauge theory (and perhaps ultimately, QCD) at
finite temperature and density, which is of great experimental
interest. Further work on this is of great interest.

Kerr--AdS black holes were studied for the first time in the AdS/CFT
context in ref.\cite{hawkingtwo}, where the authors studied three,
four and five dimensional Kerr--AdS black holes and some features of
their dual field theories. Several important contributions have been
subsequently presented on Kerr--AdS black holes
\cite{berman}--
\cite{horowitz}, addressing the subject of AdS/CFT duality.

Asymptotically AdS black holes enjoy many interesting properties
distinct from black holes which are asymptotically Minkowskian. For
example, the presence of a negative cosmological constant,
$\Lambda<0$, allows more geometries for black holes horizons than in
the $\Lambda=0$ case.  These horizons can be spherical, planar and
hyperbolic surfaces.  Another interesting property that is not shared
with $\Lambda=0$ solutions is that AdS black holes are
thermodynamically stable, in the canonical ensemble, for certain
ranges of their parameters. Schawrzchild--AdS black holes which are
large compared with the scale, $l\sim|\Lambda|^{-1/2}$, set by the
cosmological constant, have positive specific heat, which is roughly
attributable\cite{hawkpage} to the fact that AdS effectively acts as a
box with reflecting walls supplied by its natural boundary at
infinity.  Kerr--AdS black holes are also stable in this way, and
again this is not the case for ordinary Kerr. Kerr--AdS solutions have
Killing vectors which are rotating with respect to time--translation
Killing vectors, and time--like everywhere. This allows thermal
radiation to rotate with the black hole's angular velocity all the way
to infinity.  As a result, superradiance is not allowed, and the black
hole is in thermal equilibrium with the thermal radiation around it
\cite{hawkingtwo,Hawkingthree}.

In this paper we present the results of continued studies of Kerr--AdS
black holes, using the counterterm subtraction method recalled in the
next section.  We calculate stress--energy tensors as well as
gravitational actions of Kerr--AdS$_{n+1}$ spaces, for $n=4, 5$ and
$6$, where the five dimensional case is for the general solution with
two rotation parameters, while we consider only the one--parameter
solutions for higher dimensions.

In the general Kerr--AdS$_5$ case we find that the stress--energy
tensor, for zero mass parameter, exactly matches that of the weakly
coupled dual field theory (four dimensional large $N$ $SU(N)$ gauge
theory on a rotating spacetime --see below) at zero temperature.
Consequently, the Casimir energies and conformal anomalies of the dual
theory exactly match their Kerr--AdS$_5$ counterparts. This happens as
a consequence of a non--renormalization theorem which protects anomaly
coefficients from higher loop corrections, specifically for this dual
field theory. In Kerr-AdS$_7$ the situation is a little different,
since there is no non--renormalization theorem that protects
``type-A'' anomaly coefficients (reviewed below) from higher loop
corrections.  As a result, we find that the conformal anomalies on
both sides of the correspondence (the dual field theory is the six
dimensional ``$(0,2)$'' tensor multiplet field theory at large $N$)
are not exactly equal, (see ref.\cite{tseytlin} for the field theory
result) there is a factor of 4/7 difference between them, presumably
corresponding to a renormalisation in going from weak to strong
coupling. We believe that the same factor will show up in comparing a
perturbative Casimir energy calculation to the dual energy computed
for Kerr--AdS$_7$. We compute and display the latter here, although
there is no field theory calculation on the market to compare to yet.
In Kerr--AdS$_5$ and Kerr-AdS$_7$ cases we find, by examining the form
of the trace of the energy--momentum tenor, that the rotation
parameters break conformal but not scale invariance. We also argue
that at high temperature, the dual stress--energy tensors are equal up
to the usual 3/4 factor encountered in comparing such unprotected
things across the weak/strong coupling divide\cite{gubpeetkleb}.

We completed these calculations and announced them in
ref.\cite{paper2}, where we used the stress tensors as examples of the
scale/conformal invariance issue mentioned in the previous paragraph.
We suppressed some of the computational details there, promising to
display them later. This is the main purpose of this paper. We have
learned that a recent paper has appeared with computations which
overlap with some of those presented here\cite{mann2}.

\section{Computing the Action and Stress Tensor}
\label{counter}
The gravity  action on a region of spacetime ${\cal M}$, with boundary
$\partial {\cal M}$ has the form\cite{Gibbons}, 
\begin{eqnarray}
I_{\rm bulk}+I_{\rm
surf}= -{1 \over 16 \pi G}\int_{\cal M} d^{n+1}x \sqrt{-g}\left(R+{n(n-1)
\over l^2}\right)-{1 \over 8 \pi G } \int_{\partial
{\cal M}} d^{n}x \sqrt{-h} \,K.  
\end{eqnarray} 
The first term is the Einstein--Hilbert action with negative
cosmological constant $\Lambda$, which defines a natural length scale
as follows: $\Lambda{=}{-n(n{-}1)/2l^2}$. The second term is the
Gibbons--Hawking boundary term.  Here, $h_{ab}$ is the boundary metric
and $K$ is the trace of the extrinsic curvature $K^{ab}$ of the
boundary. We of course wish to consider the case where ${\cal M}$ is a
complete spacetime which is asymptotically AdS, therefore having
infinite volume. 

To deal with the divergences which appear in the gravitational action
(arising from integrating over the infinite volume), we shall use the
``counterterm subtraction''
method\cite{Henningson}--
\cite{Balasubramanian}.  The method
regulates the action by the addition of certain boundary counterterms which
depend upon the geometrical properties of the boundary of the spacetime. They
are chosen to diverge at the boundary in such a way as to cancel the bulk
divergences\cite{Henningson}--
\cite{counterterm} (see
also \cite{lau,mann1,kraus,solo}): \begin{eqnarray} I_{\rm ct}={1 \over 8 \pi
G} \int_{\partial {\cal M}}d^{n}x\sqrt{-h}\Biggl[ \frac{(n-1)}{ l}-{l{\cal R}
\over 2(n-2)}+{l^3\over2(n-4) (n-2)^2}\left({\cal R}_{ab}{\cal R}^{ab}-{n\over
4(n-1)}{\cal R}^2\right)\Bigg]\ . \label{theterms} \end{eqnarray} Here ${\cal
R}$ and ${\cal R}_{ab}$ are the Ricci scalar and tensor for the boundary
metric $h$.  Using these counterterms one can construct a divergence--free
stress--energy tensor from the total action $I{=}I_{\rm bulk}{+}I_{\rm
surf}{+}I_{\rm ct}$ by defining (see {\it e.g.}  \cite{Brown}): 
\begin{eqnarray}  T^{ab}&=& {2 \over \sqrt{-h}} {\delta I
\over \delta h_{ab}}\ .
\label{stressone}
\end{eqnarray}
The metric restricted to the boundary, $h_{ab}$, diverges
due to an infinite conformal factor, which will turn out to be
$r^2/l^2$ in coordinates we will introduce later. 

We will be interested in comparing quantities computed for ${\cal M}$
to a strongly coupled dual field theory, which resides on a spacetime
with a metric conformal to $h_{ab}$.\footnote{Sometimes this is just
  referred to as ``the theory on the boundary'', which is only roughly
  true.}  We take the background metric upon which the dual field
theory resides as
\begin{equation}
\gamma_{ab}=\lim_{r\to\infty}{l^2\over r^2}h_{ab}\ .
\label{newmetric}
\end{equation}
and so the dual strongly coupled field theory's stress--tensor,
${\widehat T}^{ab}$ \footnote{The expression for ${\widehat T}^{ab}$ has been computed for arbitrary
background manifold in ref.\cite{skenderis}.}, is related to the one in
equation~(\ref{stressone}) by the rescaling\cite{robstress}: \begin{equation}
 \sqrt{-\gamma}\,\gamma_{ab}{\widehat
T}^{bc}=\lim_{r\to\infty}\sqrt{-h}\,h_{ab}T^{bc}\ .
\label{newstress}
\end{equation}
This amounts to multiplying all expressions for $T^{ab}$ displayed later by $(r/l)^{n-2}$ before taking the limit
$r{\to}\infty$.

 \section{The General Kerr--AdS$_{5}$ Solution}
 
 The rotation group in $n+1$ dimensions is $SO(n)$. The number of
 independent rotation parameters for a localized object is equal to
 the number of Casimir operators, which is the integer part of $n/2$.
 This means that general Kerr--AdS$_5$ has two parameters.  Its metric
 is given by~\cite{hawkingtwo}:
\begin{eqnarray} 
ds^2&=&-{\Delta_{r} \over{\rho}^2}
\left(dt-{a\sin^2{\theta}\over\Xi_a}d\phi-{b
\cos^2{\theta}
\over\Xi_b}d\psi\right)^2+{\rho^2\over\Delta_\theta}
d\theta^2+{\Delta_{\theta}\sin^2{\theta}\over\rho^2}\left(adt-{(r^2+a^2)
 \over {\Xi}_a} d\phi\right)^2\nonumber\\      
&
&+{(1+r^2/l^2)\over r^2\rho^2}
\left(abdt-{b(r^2+a^2)
\sin^2\theta\over\Xi_a}d\phi-{a(r^2+b^2)
\cos^2\theta\over\Xi_b} d\psi\right)^2\nonumber\\
&
&+{\rho^2\over\Delta_{r}}dr^2+{\Delta_{\theta}\cos^2{\theta}
\over\rho^2}\left(bdt-{(r^2+b^2)\over\Xi_b}d\psi\right)^2 \ ,  \end{eqnarray} 
where
\begin{eqnarray} 
\rho&=&r^2+a^2\cos^2\theta+b^2\sin^2\theta,\nonumber\\
 \quad \Xi_a&=&1-a^2/l^2,
\quad \Xi_b=1-b^2/l^2\nonumber\\  
\Delta_{r}&=&{1\over r^2}(r^2+a^2)(r^2+b^2)(1+r^2/l^2)-2MG, \nonumber\\
\Delta_{\theta}&=&1-a^2/l^2\cos^2\theta-b^2/l^2\sin^2\theta\ .  
\end{eqnarray} 
The inverse temperature, computed by requiring regularity of the
Euclidean section, is:
\begin{equation} \beta={2\pi
    r_{+}({r_{+}}^2+a^2)({r_{+}}^2
    +b^2)l^2\over 2r_{+}^6+r_{+}^4(l^2+b^2+a^2)-a^2b^2 l^2}\ .
\end{equation} 
The stress energy tensor components calculated for general
Kerr--AdS$_5$ are considerably long expressions.  We found that once
we performed the procedures outlined in the previous section, the
resulting stress--energy tensor, proposed to be that for the
strongly coupled dual field theory, can be written in the following
compact form:
\begin{equation}
{{\widehat T}_a}^{b}={M \over 8 \pi
  l^3}\left[4u_{a}u^{b}+{{\delta}_a}^{b}\right]-{l^3 \over 64 \pi
  G}\left[ {1 \over 12} {{\delta}_a}^b R^2 -R^{cd} {R_{cad}}^b\right],
\label{compact}
\end{equation} where $R$, $R_{ab}$ and $R_{abcd}$ are the Ricci scalar, Ricci
tensor and Riemann tensor of the background metric $\gamma_{ab}$.
Here, $u^a=(1,0,0,0)$ is a unit time--like vector and $M$ is the mass
parameter. We shall discuss in more detail the above form of the
stress--energy tensor by the end of the subsection~\ref{complete},
where there and in the next subsection we test whether this is indeed
related to the stress tensor of the large $N$ $SU(N)$ Yang--Mills
theory.

The mass and angular momenta calculated from this tensor are given by:
\begin{eqnarray}
{\cal M}&=&{\pi l^2\over 96G\Xi_a\Xi_b}[7\Xi_a\Xi_b+{\Xi_a}^2+{\Xi_b}^2+72 G
M/l^2]\ ,
\end{eqnarray}
and
\begin{equation} 
{\cal J_{\phi}}={\pi M a \over 2 {\Xi_a}^2 \Xi_b},\quad   {\cal
J_{\psi}}={\pi M b \over 2 {\Xi_b}^2 \Xi_a}\ .  
\label{angular}
\end{equation} 
The angular velocities on the horizon have the form: 
\begin{equation} 
{\Omega}_{\phi}= a {{\Xi}_a \over r_{+}^2+a^2},\quad
{\Omega}_{\psi}= b {{\Xi}_b \over r_{+}^2+b^2}\ .
\end{equation}  
The action is given by
\begin{eqnarray}
I_5&=&-{\pi \beta l^2 \over 96 
\Xi_a\Xi_b G}\left[12({r_+}^2/l^2)(1-\Xi_a-\Xi_b)+{\Xi_a}^2+{\Xi_b}^2
+\Xi_b\Xi_a+12{r_+}^4/l^4-2(a^4+b^4)/l^4\right.\nonumber\\
   & &\hskip2cm\left.-12(a^2b^2/l^4)({r_+}^{-2}l^{-2}-1/3)-12 \right] \ .
\end{eqnarray} 
while the area of the horizon is 
\begin{equation} 
{\cal A}={2\pi^2G ({r_+}^2+a^2)({r_+}^2+b^2) \over {r_+}\Xi_a \Xi_b}\ .
\end{equation} 

We note that the above quantities satisfy the following
thermodynamical relation for $n=4$
\begin{eqnarray}
S=\beta\left( {\cal M}-{\Omega}_i{\cal J}_i\right)-I_{n+1}={{\cal A}\over
4G}\ , \label{thermo}
\end{eqnarray}
which is a highly non--trivial check of our calculations. (Note that
the expressions for angular momenta (\ref{angular}) are different from
those given in ref.\cite{hawkingtwo}, where there is a missing factor of
${\Xi_a}^{-1}$ or ${\Xi_b}^{-1}$ in their expressions. The angular
momenta and the other quantities calculated in that reference
therefore do not satisfy the above thermodynamical relation, but with
the expressions presented here, they do.)

\subsection{The Conformal Anomaly} 
The dual field theory is proposed to be strongly coupled ${\cal N}=4$
supersymmetric $SU(N)$ Yang--Mills theory with the supersymmetry
broken by the rotation and the finite temperature (mass of the black
hole).  The spacetime metric on which the dual field theory resides is
found, using the definitions in section~\ref{counter}, to be:
\begin{equation} 
ds^2=
\left[-dt^2+{2a\sin^2{\theta}\over\Xi_a}dtd\phi
+{2b\cos^2{\theta}\over\Xi_b}dtd\psi+l^2{d\theta^2
  \over\Delta_{\theta}}
+l^2{\sin^2{\theta} \over \Xi_a}d\phi^2+{l^2\cos^2
\over \Xi_b}{\theta}d {\psi}^2\right]\ .
\end{equation}
It is important to note that this generalized rotating Einstein
universe\cite{hawkingtwo} is conformally flat. This will be of great
use later.

Now we would like to compare some of the results we have found on the
AdS side of the correspondence with their dual on the field theory
side.  It is difficult to calculate physical quantities in the
strongly coupled field theory in order to make the comparison
directly. Sometimes, however, there is a non--renormalization theorem
which protects anomaly coefficients from higher loop corrections.  

When this is true, the Casimir energy and the anomaly coefficients for
both strongly coupled and weakly coupled limits of the theory should
be the same.  These anomaly coefficients are protected because the
trace of the stress--energy tensor and the divergence of the $SU(4)$
R--symmetry current are in the same supermultiplet, which is called
anomaly multiplet.  The divergence of the R--current is just the
chiral anomaly, which is one loop exact by Adler--Bardeen theorem.
(See {\it e.g.} refs.\cite{nonrenorm} for  relevant
discussions.)

Unfortunately, our treatment does not include any supersymmetry, as it
is broken by the rotation, and therefore we should not expect that we
have any right to protection from a non--renormalisation theorem.
Fortunately, higher loop corrections (corrections beyond planar
diagrams) are of order $O(1)$ and $O({1 \over N})$
\cite{anselmi}--
\cite{bilal} and so will not change the one loop
results as long as we study the large $N$ limit, which is precisely where the
duality holds (at least if we stay in pure supergravity). This is the
main reason that the conformal anomaly and Casimir energy for weakly
coupled fields will be exactly equal to their duals on the gravity
side, we will see in the following.

The definition of the stress tensor for the dual field theory that we
shall use for weak coupling is the one introduced by Brown and Cassidy
\cite{browncassidy}. They calculated the renormalized stress tensor
for free fields (conformally coupled scalar, Weyl spinor and $U(1)$
gauge field). Their expression is given by:
\begin{eqnarray}
  <\!{{\widehat T}^{s}}_{ab}\!>={H_{ab}}^{(4)}-{1\over 16
    \pi^2}\left[{1\over 9} \alpha^{s} {H_{ab}}^{(1)}+ 2 \delta^s
    {H_{ab}}^{(3)}\right]\ , 
\label{stressed}
\end{eqnarray} 
where ${H_a^b}^{(1)}$, ${H_a^b}^{(3)}$, $\alpha^s$ and $\delta^s$ are
defined in ref.\cite{BD}.  (Note that they use spacetime indices
$(\mu,\nu)$ for the field theory and $\beta^s$ for the coefficient of
${H_a^b}^{(3)}$ while here we use $(a,b)$ for the indices and $\delta^s$ for
the coefficient.) If the expectation value is over vacuum state,
${H_{ab}}^{(4)}$ will be zero, since it is identified with the vev of the
stress--energy tensor in Minkowski space. We will use this fact in section
\ref{complete}.  The label $s{\in}\{0,1/2,1\}$ distinguishes the spin of the
field for which the labelled coefficients $\alpha^s$ and $\delta^s$ are
computed. We can now compare this to the the non--thermal ({\it i.e.,} the
$M$--independent) part of the tensor which we have computed on the gravity
side, inserting the spins appropriate to the content of the dual theory: a
gauge field, six scalars and four Weyl spinors.

According to the AdS/CFT  relation, one should define the
Casimir energy for the field theory dual to the Kerr--AdS spacetime as
the contribution to the total energy of the spacetime which is
independent of the black hole's mass\cite{Balasubramanian}. We find
that the two energies exactly match  and they are given by:
\begin{equation} 
{\cal E}={N^2  \over 48 l\Xi_a \Xi_b}(7\Xi_a\Xi_b+{\Xi_a}^2+{\Xi_b}^2)\ .
\label{energy1}
\end{equation} where we used the standard entry in the AdS/CFT
dictionary\cite{maldacena}
\begin{equation} 
{1 \over G}={2N^2 \over \pi l^3}.  
\label{diction}
\end{equation} 
This Casimir energy agrees with the one calculated in ref.\cite{paper}
for one parameter Kerr--AdS$_5$ case as $b{\rightarrow} 0$.

Furthermore the two stress--energy tensors at zero mass parameter
limit, are equal.  This is not surprising, since conformal anomaly
fixes the whole stress tensor in conformally flat background, once the
quantum state is determined\cite{BD}.

Let us now investigate conformal anomaly for dual field
theory,\cite{witten, Henningson} on the boundary.  The general form of
Weyl anomaly in arbitrary dimension $n$ is given by (see {\it e.g.}
the discussions in refs.\cite{deser,tseytlin}):
\begin{equation} 
{T_a}^a= c_0 E_n + \Sigma c_i I_i + \nabla_i J^i
\label{anom}
\end{equation} 
where $E_n$ is Euler density in $n$ dimensions, $I_i$ are
invariants which contain the Weyl tensor and its derivatives, and the
last term is a collection of total derivative terms that can be
removed by adding suitable local couterterms.\footnote{This term is
  the generalization of the familiar $\Box R$ in four dimensions.}
The first type of term is called ``type~A'', the second ``type~B'',
and the last ``type~D''.  It is important to note that the
coefficients of all terms are regularisation scheme independent {\sl
  except} the type~D anomaly.

In the two non--trivial cases of spherical Kerr--AdS$_5$ and
Kerr--AdS$_7$, the boundary metric is conformally flat which means
that the Weyl invariant terms vanish. The only surviving non--trivial
term is the Euler density which is locally a total derivative. The
Euler density integral is a topological invariant which vanishes for
the boundaries of Kerr--AdS$_5$ and Kerr--AdS$_7$ spaces.  In {\it
  special cases} the Euler density is proportional to a combination of
terms in in the type D anomaly.  This doesn't make the type A anomaly a type
D anomaly, since the coefficient $c_0$ is scheme
independent. Attempting to add
counterterms to remove the type A anomaly will change the anomaly
coefficient $c_0$, which means that we have moved to  a different
theory, instead of merely changing scheme.

We find\cite{paper2} that this special situation is just what happens
for the Kerr--AdS$_5$ and Kerr--AdS$_7$ solutions.  The conformal
anomaly for two parameters Kerr--AdS$_5$ has the following form:
\begin{eqnarray}
{T_a}^a=-{(a^2-b^2) N^2 \over 4 \pi^2
l^6}\left[a^2/l^2\cos^2\theta(3\cos^2\theta-2)
+b^2/l^2(\cos^2\theta(-3\cos^2\theta+4)-1)-\cos 2 \theta\right]\ ,
\end{eqnarray}
which matches exactly the general form of the trace anomaly in four
dimensions.  In terms of the Euler density, it can be written simply
as
\begin{equation} 
{T_a}^a= c_0 E_n 
\end{equation} 
where $c_0=-{N^2/\pi^2}$. 
The integrated anomaly is zero as in the one--parameter case
\begin{equation} 
\int dx^4 \sqrt{-\gamma} {T_a}^a=0\ .
\end{equation} 
Since the energy--momentum tensor is traceful, but the trace is an
irremovable total derivative, we see that for arbitrary rotation
parameters we have broken conformal invariance but preserved scale
invariance. This is discussed at length in ref.\cite{paper2}. It is
interesting to note that when the rotation parameters are equal, the
anomaly vanishes identically.

\subsection{The Stress--Energy Tensor at High Temperature} 
\label{complete}
We show in this section that at high temperature the stress--energy
tensor we computed for the general Kerr-AdS$_5$ case matches that of
the dual field theory, up to the usual 3/4 factor encountered in going
from weak to strong coupling.

Let us start with the renormalized stress--energy tensor of Brown and
Cassidy for the field theory given in equation (\ref{stressed}).
Replacing the vacuum expectation value by the ensemble average $<\!
\,\,\,\!  >_{\beta}$ one obtains the thermal stress--energy tensor:
\begin{eqnarray}
<{{T}_{a}}^b (\gamma)>_{\beta}&=&{\omega}^{-4}<\!
{{T}_{a}}^b(\eta)\!>_{\beta}-{1\over 16 \pi^2}\left[{1\over 9} \alpha
  {{H^{(1)}}_a}^b+ 2 \delta {{H^{(3)}}_a}^b \right]. 
\label{hotstressed}
\end{eqnarray} 
Here $\gamma_{ab}$ is the spacetime metric of the theory, $\eta_{ab}$ is
that of Minkowski spacetime, and $\omega$ is the conformal factor
relating the two metrics ({\it i.e.}, $\gamma_{ab}={\omega}^2 {\eta}_{ab}$).
The  boundary metric ({\it i.e.}, rotating Einstein universe) is conformal to
Minkowski space, since rotating Einstein universe
after a change of variables is conformal to Einstein universe\cite{hawkingtwo,cassidy}, which is known to be conformal to Minkowski
space.  Notice that upon thermalizing the Brown--Cassidy expression, the
coefficients $\alpha$ and $\delta$ will not depend on $\beta$ ({\it   i.e.},
the temperature), since the same coefficients appear in the trace which is
non--thermal.\footnote{The reason is that, the boundary   metric does not
depend on the mass parameter since the mass term   vanishes at the boundary. 
But trace anomaly is a local geometric   expression which depends only on the
metric and its derivatives,   hence it will not depend on $\beta$.}  Notice
that the first term on the right hand side of the above expression vanishes at
zero temperature if the manifold is conformally related to the whole
Minkowiski space\cite{BD}.

One way to understand the above expression is to think of it as the
quantum version of the classical stress-energy tensor transformation
law under conformal transformation $ \gamma_{ab}={\omega}^2
{\eta}_{ab}$. If a classical action is invariant under conformal
transformation, then the stress-energy tensor will transform as
\begin{equation} {{ T}_{a}}^b\!( \gamma)={\omega}^{-4}{{
      T}_{a}}^b(\eta)
\end{equation} provided that ${T_a}^b(\eta)$ is conserved. The existence
of the last two terms in the right hand side of the expression is due
to pure quantum effects (one loop corrections). This follows from
realizing that they are the source of both the Casimir energy and the
trace anomalies, which are pure quantum effects.

In ref.\cite{paper} we argued that in the limit $M \rightarrow 0$ the
two stress--energy tensors on both sides of the duality are identical.
The reason is that the field theory stress-energy tensor is protected
from higher loop corrections as we mentioned earlier. Furthermore, and
as we also already mentioned, once we have matched the conformal
anomaly, it determines the entire tensor (at zero temperature) in the case
we have here that the spacetime the field theory is defined on is
conformally flat\cite{BD}.

Now let us try to probe another aspect of this duality by going to
another limit ---the high temperature limit--- checking the relation
between the stress-energy tensors at weak and strong coupling (the
latter defined by the AdS computation).  At high $T=\beta^{-1}$, we
can ignore quantum corrections in the field theory, since these quantum
corrections do not depend on temperature, while the Minkowiski
stress-energy tensor (the first term in the right hand side of
equation (\ref{hotstressed})) goes as $T^4$. So in this limit we have:
\begin{equation} 
<\!{{ T}_{a}}^b\!(\gamma)>_{\rm(FT)}={\omega}^{-4}<\!{{T}_{a}}^b(\eta)\!> 
\end{equation} 
The previous relation is just the transformation law of stress-energy
tensor under conformal transformation for a conformally invariant
effective action.  $<\!{{ T}_{a}}^b(\eta)\!>$ is traceless and has the
form
\begin{equation} <\!{{ T}_{a}}^b(\eta)\!>={{\rho}^{\rm M}\over 3}\left[
    4u_a u^b +{{\delta}_a}^b\right]
\end{equation} where the
    density is:
\begin{equation} 
\rho^{\rm M}={\pi^2 N^2 \over 2 {{\beta}_{\rm M}}^4}\ ,
\end{equation} 
(the label ``M'', is for ``Minkowski''), and again, $u_a$ is a
time--like unit vector.  The stress--energy tensor of our field theory
is given by
\begin{equation} <\!{{
      T}_{a}}^b(\gamma)\!>_{\rm (FT)}={{\rho}^{\gamma}\over 3}\left[ 4u_a u^b
    +{{\delta}_a}^b\right]\ ,\quad\mbox{where}\quad \rho^{\gamma}={\pi^2 N^2
    \over 2 {{\beta}_{\gamma}}^4}\ ,
\end{equation} 
with the label ``$\gamma$'' for the spacetime of our field theory.  Notice
that density gets scaled under conformal transformation
\begin{equation} 
\rho^{\gamma}=\omega^4 \rho^M.
\end{equation} 
Then the field theory stress--energy tensor after dropping the label $\gamma$
is given by
\begin{equation} <\!{{
      T}_{a}}^b(\gamma)\!>_{\rm (FT)}={\pi^2 N^2
    \over 6 {{\beta}}^4}\left[ 4u_a u^b
    +{{\delta}_a}^b\right]\ .
\end{equation} 
On the gravity side one can express the mass parameter of
Kerr--AdS$_5$ black hole as a function of $\beta$.  For small $\beta$
we get
\begin{equation} 
M={\pi^4 l^6 \over 2 G\beta^4}\ .
\end{equation} 
Substituting this value of $M$ in the expression of stress--energy
tensor in equation~(\ref{compact}), and using
equation~(\ref{diction}), the stress--energy tensor on the gravity
side is
 \begin{equation} <\!{{
    T}_{a}}^b(\gamma)\!>_{\rm (AdS)}={\pi^2 N^2 \over 8 {\beta}^4}\left[ 4u_a
u^b   +{{\delta}_a}^b\right] \end{equation} 
The two expressions for the stress-energy tensors are the same up to the
usual~3/4 factor, which is now familiar ratio between energetic quantities
from  the strong and weak coupling regime of the theory\cite{gubpeetkleb}.

The above argument can be used for rotating solutions with charge, as
well with spherical or flat horizons, since after a change of
coordinates, the boundary metric for these cases are also conformal to
the whole Minkowiski space. In these cases we have temperature
dependence similar to the Kerr--AdS case, because the mass as a
function of temperature, contains other terms that depend on charges
but which can be ignored at high temperature. The boundary metric of
rotating charged black holes will be the same as the one for rotating
black hole case, since terms that contain charges will vanish at the
boundary. We expect that one can do the same for black holes with
hyperbolic horizons. The only difference is that hyperbolic black
holes boundaries are conformal to Rindler space, but this can be
chosen as a part of Minkowiski space.

\section{The Kerr--AdS$_{6}$  Solution}
The general Kerr--AdS$_6$ has two rotation parameters, but we will
consider only one parameter solution. The metric for 
Kerr--AdS$_{6}$ with one rotation parameter is given by
\begin{eqnarray} ds^2&=&-{\Delta_{r} \over {\rho}^2}\left(dt-{a
      \sin^2{\theta} \over \Xi } d\phi\right)^2 +r^2 \cos^2{\theta}
  d\psi^2+{{\rho}^2\over\Delta_\theta}d\theta^2+{\rho^2 \over
    \Delta_{r}}dr^2\nonumber\\
  & &+{\Delta_{\theta}\sin^2{\theta}\over\rho^2}\left(adt-{(r^2+a^2)
      \over \Xi} d\phi\right)^2+r^2 \cos^2{\theta} d \Omega_{2}^2,
\end{eqnarray} where 
\begin{eqnarray}
\rho &=& r^2+a^2cos^2(\theta)\ ,\quad \Xi=1-a^2/l^2,\nonumber\\
     \Delta_{r}&=&(r^2+a^2)(1+r^2/l^2)-2MG/r\ ,\nonumber\\
\Delta_{\theta}&=&1-a^2/l^2 \cos^2(\theta).
\end{eqnarray}
$d \Omega_{2}^2$ is the metric on the two--sphere
\begin{equation}
d\Omega_2^2=d\psi^2+\sin^2\psi d\eta^2.
\end{equation} 
The inverse temperature is given by
\begin{equation} 
\beta={4 \pi (r_{+}^2+a^2)r_{+} \over
5r_{+}^4/l^2+3r_{+}^2(1+a^2/l^2)+a^2}\ ,
\end{equation} 
the angular velocity has the form 
\begin{equation}
\Omega=a {\Xi \over r_{+}^2+a^2}\ .
\end{equation}
Here, $r_+$ is the location of the horizon, the largest root of
$\Delta_r$.

The non--vanishing components of the
stress--energy tensor at large $r$ are given by,
\begin{eqnarray}
T_{tt}&=&{M \over lG 2\pi r^3}+O\left({1\over r^4}\right)\ 
,\nonumber\\
        T_{t\phi}&=&-{Ma \sin^2{\theta} \over lG 2\pi \Xi r^3}+
O\left({1\over 
r^4}\right)\ ,\nonumber\\
     T_{\phi\phi}&=&{Ml \over {\Xi}^2 8\pi G r^3}(4 \Xi-5 
\Delta_{\theta})+O\left({1\over r^4}\right)\ ,\nonumber\\
 T_{\theta\theta}&=& {Ml \over {\Delta}_{\theta}8\pi G r^3}+ O\left({1\over
r^4}\right)\ ,\nonumber\\
     T_{\psi\psi}&=&{l\cos^2{\theta}M \over 8\pi G r^3}+O\left({1\over 
r^4}\right)\,\nonumber\\
     T_{\eta\eta}&=& {\sin^2{\theta} \cos^2{\theta} lM \over 8\pi G  
r^3}+O\left({1\over
r^4}\right)\ 
\end{eqnarray}
Using the definition for a conserved charge, one
can calculate the mass and angular momentum of the solution:
\begin{eqnarray}
{\cal M}={4 \pi \over 3 \Xi}M\ ,\quad
{\cal J}={2 \pi \over 3 \Xi^2}aM\ .
\label{energy2}
\end{eqnarray}
Also the action is
\begin{eqnarray}
I_{6}&=&-{4\pi^2(r_{+}^{2}+a^{2})r_{+}\left[r_{+}^5/l^2+r_{+}^3/l^2a^2-MG
\right] 
\over \Xi G\left(
5r_{+}^4/l^2+3r_{+}^2(1+a^2/l^2)+a^2\right)}\ .
\label{theaction1}
\end{eqnarray}
The area of the horizon is given by
\begin{eqnarray}
{\cal A}=8 \pi^2G {r_{+}^2(r_{+}^2+a^2) \over 3\Xi}\ .
\end{eqnarray}
The above quantities satisfy the thermodynamical relation (\ref{thermo}) for
$n=5$.

The boundary metric is given by
\begin{equation} ds^2=
\left[-dt^2+{2a\sin^2{\theta}\over\Xi}dtd\phi+l^2{d\theta^2 \over
\Delta_{\theta}}+l^2{\sin^2{\theta} \over \Xi}d\phi^2+l^2\cos^2{\theta}
d {\Omega_{2}}^2\right]\ .
\end{equation}

Defining the tress tensor of the dual field theory by removing the
conformal factor, one can put it in the compact form:
\begin{equation} 
T_{a b}={M \over 2\pi l^4}\left[5u_au_b+\gamma_{ab}\right]
\end{equation} 
where $u^a=(1,0,0,0,0)$ and $\gamma_{ab}$ is the boundary metric. It
is interesting to notice that the tensor can be put in the standard
form even in the presence of non--vanishing rotation parameter. We
noticed this for Kerr--AdS$_4$ in ref.\cite{paper}. Conformal
invariance together with conservation of the stress--energy tensor
seems to constrain the form of the tensor. This property is likely
shared by all odd dimensional field theory stress--energy tensor of
this sort.

\section{The Kerr--AdS$_{7}$ Solution} 
The general solution for Kerr--AdS$_{7}$ contains three rotation
parameters, but we are going to consider only one parameter solution.
The metric for the Kerr--AdS$_{7}$ is \begin{eqnarray}
  ds^2&=&-{\Delta_{r} \over \rho^2}\left(dt-{a \sin^2{\theta} \over
      \Xi } d\phi\right)^2 +r^2 \cos^2{\theta}
  d\psi^2+{{\rho}^2\over\Delta_\theta}d\theta^2+{\rho^2 \over
    \Delta_{r}}dr^2+\nonumber\\
  & &{\Delta_{\theta}\sin^2{\theta}\over\rho^2}\left(adt-{(r^2+a^2)
      \over \Xi} d\phi\right)^2+r^2 \cos^2{\theta} d \Omega_{3}^2,
\end{eqnarray}
where the functions in the metric are the same except
\begin{equation} 
\Delta_{r}=(r^2+a^2)(1+r^2/l^2)-2MG/r^2\ ,
\end{equation} 
and $d \Omega_{3}^2$ is the metric on a three--sphere
\begin{equation}
d\Omega_3^2=d\psi^2+\sin^2\psi d\eta^2 +\cos^2 \psi d \beta^2\ .
\end{equation}   
The inverse temperature is
\begin{equation} 
\beta={2 \pi (r_{+}^2+a^2)r_{+} \over 3r_{+}^4/l^2+2r_{+}^2(1+a^2/l^2)+a^2}\ ,
\end{equation} 
and the angular velocity is the same as in the previous section.  Once
again $r_{+}$ is the location of the horizon, the largest root of
$\Delta_r$.  The non--vanishing components for the stress--energy
tensor at large $r$ are given by,
\begin{eqnarray}
T_{tt}&=&{l^3 \over 640 \pi G r^4}\left[
(1+a^2/l^2)(-131a^2/l^2-423(a^4/l^4)\cos^4{\theta})+219
(1+a^4/l^4)(a^{2}/l^{2})\cos^2{\theta}\right.\nonumber\\                     
            & &\left.+25(1+a^6/l^6)+235
(a^{6}/l^6)\cos^6\theta+400MG/l^4+492(a^4/l^4)
\cos^2{\theta}\right]+O({r^{-5}})\ ,\nonumber\\         
T_{t\phi}&=&-{l^3 a \sin^2\theta \over 640 \pi G
r^4\Xi}\left[-231(a^{6}/l^6)\cos^4\theta+5+a^2/l^2
+55(a^{6}/l^6)\cos^6\theta-101a^{4}/l^{4}+400MG/l^4\right.\nonumber\\
&
&\left.+189(a^{6}/l^6)\cos^2{\theta}-25a^6/l^{6}
+168(a^{4}/l^4)\cos^2{\theta}-51(a^{4}/l^4)\cos^4\theta
-3(a^{2}/l^2)\cos^2{\theta}\right]+O({r^{-5}}),\nonumber\\                     
T_{\phi\phi}&=&{al^5\sin^2\theta \over 640 \pi G r^4
\Xi^2}\left[102a^{4}/l^{4}+51(a^{4}/l^4)\cos^4{\theta}
-55(a^{6}/l^6)\cos^6\theta \Xi-171(a^{4}/l^4) 
\cos^2\theta+80MG/l^4\right.\nonumber\\                              
&
&+189(a^{8}/l^8)\cos^2\theta+3(a^{2}/l^2)\cos^2{\theta}
+4a^{2}/l^{2}-231(a^8/l^8)\cos^4{\theta}+180(a^{6}/l^6)\cos^4{\theta}
-5\nonumber\\                  
&
&\left.+480MG(a^2/l^6)\cos^2{\theta}-21(a^6/l^6)\cos^2\theta
-76a^6/l^6-25a^8/l^8+400(a^2/l^6)MG\right]+ O{r^{-5}}),\nonumber\\    
            T_{\theta\theta}&=&-{l^5 \over 640 \pi
G\Delta_{\theta}r^4}\left[5(a^2/l^2)\Xi-80MG/l^4
+3(a^2/l^2)\cos^2{\theta}(1+a^4/l^4)\right.\nonumber\\            
      &&\left.+66(a^4/l^4)\cos^2{\theta}-45(a^4/l^4)
\cos^4{\theta}(1+a^2/l^2)+25(a^6/l^6)\cos^6\right]+O({r^{-5}})\
,\nonumber\\
       T_{\psi\psi}&=&-{l^5\cos^2{\theta} \over 640 \pi G r^4}\left[5
\Xi(1-a^{4}/l^4)-80MG/l^4-51(a^{4}/l^4)\cos^4{\theta}(1+a^2/l^2)\right.
\nonumber\\                  
& &\left.-3(a^2/l^2)\cos^2\theta(1+a^4/l^4)
+46(a^{4}/l^4)\cos^2{\theta}
\right]+O({r^{-5}}), \nonumber\\    
T_{\eta\eta}&=&\sin^2{\psi}T_{\psi\psi},\nonumber\\  
T_{\beta\beta}&=&\cos^2{\psi}T_{\psi\psi}.
\end{eqnarray}
The mass and angular momentum of the solution are:
\begin{equation}
{\cal M}=-{{\pi}^2 l^4\left[ a^6/l^6+5a^4/l^4+50\Xi-800MG/l^4\right]\over
1280G \Xi},\quad
{\cal J}={ \pi^2 \over 4 \Xi^2}aM\ .
\label{energy3}
\end{equation}
The energy dual to the Casimir energy of the field theory is given by
setting $M=0$ in the above, to give:
\begin{equation}
{\cal E}=-{{\pi}^2 l^4\left[ a^6/l^6+5a^4/l^4+50\Xi\right]\over
  1280\Xi G}\ .
\end{equation} 
The action is
\begin{eqnarray}
I_{7}&=&-{\pi^3(r_{+}^{2}+a^{2})r_{+}l^4\over 640\Xi G \left(
3r_{+}^4/l^2+2r_{+}^2(1+a^2/l^2)+a^2\right)}\times\nonumber\\&&\hskip3cm
\left[160 
(r_{+}^6/l^6+r_{+}^4/l^6 a^2-MG/l^4)+5a^4/l^4+50\Xi+a^6/l^6 \right]\ .
\label{theaction2}
\end{eqnarray}
The area of the horizon is
\begin{eqnarray}
{\cal A}= \pi^3 G{r_{+}^3(r_{+}^2+a^2) \over \Xi}\ .
\end{eqnarray}
The above quantities satisfy the thermodynamical relation
(\ref{thermo}) for $n=6$.

\subsection{The Conformal Anomaly}
The field theory to which the AdS$_7$ theory is supposed to be dual
is\cite{maldacena} the ``$(0,2)$'' supersymmetric tensor field theory
at large $N$. (See ref.\cite{review} for a review.)

The metric on which the dual field theory resides is given using the
procedure of section~\ref{counter} by
\begin{equation} 
ds^2=
\left[-dt^2+{2a\sin^2{\theta}\over\Xi}dtd\phi+l^2{d\theta^2 \over
\Delta_{\theta}}+l^2{\sin^2{\theta} \over \Xi}d\phi^2+l^2\cos^2{\theta}
d {\Omega_{3}}^2\right]\ ,
\end{equation}
a higher dimensional generalization of the rotating Einstein universe.
Taking the trace of the stress--energy tensor given in the previous
section yields, after taking the limits in section \ref{counter}:
\begin{eqnarray} 
{\widehat T}_a^a&=&-{a^2 N^3 \over 2\pi^3 l^8
    }\left[5a^4/l^4 \cos^6\theta -8 \cos^4\theta
    a^2/l^2(1+a^2/l^2)-2(1+a^2/l^2)\right.\nonumber\\ &
  &\left.+3\cos^2\theta(1+a^4/l^4+3a^2/l^2)\right]
\end{eqnarray}
where we used the relation\cite{maldacena} $N^3{=}{3\pi^2 l^5/ 16 G}$
between the field theory and the gravitational parameters.

Just as we saw for the case of Kerr--AdS$_5$, we see that this trace
is a total derivative. Furthermore, we note that it can be written in
terms of the Euler density:
\begin{equation}
{\widehat T}_a^a=-{N^3 \over 4508 \pi^3}E_6\ .
\end{equation}
The Euler density $E_6$ is displayed in {\it e.g.}
ref.\cite{tseytlin}, where the field theory is discussed at weak
coupling. The coefficient matches the results in
ref.\cite{Henningson,kraus,tseytlin}. Just as in the
four dimensional case, we see that for this special situation, the
Euler density can be written in terms of type D quantities. In the
notation of ref.\cite{tseytlin}, it is of the form
$\sum_{i=1}^7d_iC_i$, with $d_5$ and $d_7$ zero, since they depend on
the Weyl tensor, and
$\{d_1,d_2,d_3,d_4,d_6\}=\{0,1/9,1/72,-5/12,1/72\}$.  
(This is a
useful parameterization, but of course, not unique.)

The conformal anomaly of the free field theory has been calculated in
ref.\cite{tseytlin} and the results are similar to the anomaly we
computed on the AdS side, up to a factor of 4/7. The discrepancy
between the two results is due to the fact that the coefficient in
front of the Euler density in six dimensions is not controlled by a
non--renormalization theorem as in the four--dimensional case.
Consequently this coefficient will be different from the weakly
coupled theory which may explain the 4/7 factor.  Unfortunately there
is no Casimir energy calculation for the free theory, and so we cannot
compare it with our results. Nevertheless, we expect that the Casimir
energy for the free theory will be different from the one calculated
on the gravity side by exactly the same factor, since the conformal
anomaly and the Casimir energy depend on the same coefficient, $c_0$,
defined in equation (\ref{anom}).

\section{Concluding Remarks}
In closing, we note that overall, as stated in the introduction for
other AdS black holes, the properties of the action and stress tensor
of the Kerr--AdS solution in various dimensions seem to be consistent
with a dual interpretation as those of a field theory in one dimension
fewer. This is a quite satisfyingly successful examination of aspects
of the AdS/CFT correspondence. This work is also an extensive
demonstration of the usefulness of the counterterm subtraction method
for computing intrinsic properties of quite complicated
asymptotically AdS spacetimes.

It would be interesting to study the full thermodynamic phase
structure of these higher dimensional rotating black holes, as done in
four dimensions in ref.\cite{klemm2}. We expect that the properties of
AdS will endow these thermodynamics with the physical properties
required by the duality, in contrast to the $\Lambda=0$ case.

While the dual field theories to Kerr--AdS reside on rotating
spacetimes ---generalizations of the rotating Einstein universe---
which are situations of less immediate concern than, {\it e.g.}, gauge
field theory at high density (to which the Reissner--Nordstrom black
holes seem to be relevant), it is nevertheless an interesting arena in
which to study the AdS/CFT correspondence and its generalizations.

We expect that there is more to be learned by further study of these
systems. Particularly interesting would be further information from
the pertubrative field theory side in the Kerr--AdS$_7$ case: The
Casimir energy of the (0,2) tensor multiplet on the (rotating) six
dimensional Einstein universe is an example of a quantity which would
be nice to compare with the result we have presented here.

An obvious extension to study, for completeness, is the inclusion of
all of the rotation parameters for AdS$_6$ and AdS$_7$, and comparison
to the corresponding field theory quantities. As the two--parameter
Kerr--AdS$_5$ case and the one parameter Kerr--AdS$_7$ case were both
rather computationally intensive, we have not yet made serious
attempts to progress beyond that level of complexity. We hope that
progress on this can be made, in order to complete the story.

\section*{Acknowledgements} 
AA's work was supported by an NSF Career grant, \#PHY--9733173. 

{} 
\end{document}